\documentclass[11pt,letterpaper]{article}
\usepackage{systeme} 
\usepackage[utf8]{inputenc}
\usepackage{multirow} 
\usepackage{booktabs} 
\usepackage[english]{babel}
\usepackage{amsmath}
\usepackage{amsfonts}
\usepackage{amssymb}
\usepackage{graphicx}
\usepackage[small,bf]{caption} 
\usepackage[left=3cm,right=3cm,top=2cm,bottom=3cm]{geometry}

\author{Jorge Pinochet}
\title{\textbf{General relativity in a nutshell II}}
\begin{document}

\author{Jorge Pinochet$^{*}$\\ \\
 \small{$^{*}$\textit{Facultad de Ciencias Básicas, Departamento de Física. }}\\
  \small{\textit{Centro de Investigación en Educación (CIE-UMCE),}}\\
 \small{\textit{Núcleo Pensamiento Computacional y Educación para el Desarrollo Sostenible (NuCES).}}\\
 \small{\textit{Universidad Metropolitana de Ciencias de la Educación,}}\\
 \small{\textit{Av. José Pedro Alessandri 774, Ñuñoa, Santiago, Chile.}}\\
 \small{e-mail: jorge.pinochet@umce.cl}\\}

\date{}
\maketitle

\begin{center}\rule{0.9\textwidth}{0.1mm} \end{center}
\begin{abstract}
\noindent The aim of this work is to use the notions of Riemann's geometry introduced in Part I, to analyze the foundations of Einstein's theory of general relativity. \\ \\


\begin{center}\rule{0.9\textwidth}{0.1mm} \end{center}
\end{abstract}

\maketitle

\section{Introduction}

The aim of the second part of this work is to use the notions of Riemannian geometry introduced in the first part to study the foundations of general relativity (GR), a geometric theory of gravity in which attractive Newtonian forces are replaced with space-time curvature. The article is aimed at those who have mastered the basics of special relativity (SR) and Newtonian gravitation.

\section{The metric of curved spacetime}

In Einstein's universe, objects moving freely under the effects of gravity simply follow geodesic paths dictated by the curvature of spacetime. To develop GR, Einstein used the powerful mathematical arsenal of Riemannian geometry, converting the flat, rigid spacetime of SR into a flexible, dynamic four-dimensional manifold that bends in the presence of mass or its equivalent in energy. As required by Riemannian geometry, this four-dimensional manifold has an associated metric tensor and its corresponding metric, which contain all the information about spacetime.\\

To define the metric of curved spacetime we must generalise Eqs. (31) and (32) introduced in Part I, which describe the flat spacetime of SR. This generalisation is made by replacing the components $\eta_{\mu \nu}$ of the Minkowski metric tensor, which are constant, by the components $g_{\mu \nu}$ that are functions of the coordinates and the mass-energy distribution generated by the spacetime curvature, so that [1]:

\begin{equation} 
ds^{2} = \sum_{\mu=0}^{3} \sum_{\nu=0}^{3} g_{\mu \nu} dx_{\mu} dx_{\nu},
\end{equation}

where $ \mu, \nu = 0$ denotes the temporal coordinate, and $\mu, \nu = 1,2,3$ denote the spatial coordinates. The line element $ds$ is invariant under coordinate transformations. The set of $g_{\mu \nu}$ defines a metric tensor, which is represented as a $4 \times 4$ symmetric matrix with 16 components [2]:

\begin{equation} 
\left[ g_{\mu \nu} \right] = \begin{pmatrix} g_{00} & g_{01} & g_{02} & g_{03} \\ g_{10} & g_{11} & g_{12} & g_{13} \\ g_{20} & g_{21} & g_{22} & g_{23} \\ g_{30} & g_{31} & g_{32} & g_{33} \end{pmatrix}.
\end{equation}

However, since this matrix describes a manifold of $n = 4$ dimensions, according to Eq. (2) of Part I, the metric tensor has only $n(n+1)/2 = 10$  independent components. As we will soon see, the central problem of GR is to find these ten components for a given mass-energy distribution. But before that, it is important that we understand the concept of curvature of spacetime and its relationship with gravity.

\section{Geodesics, curvature, and gravity}

A \textit{spacetime geodesic} is by definition a locally straight line in spacetime. When a particle moves freely under the influence of gravity, with no forces acting on it, its trajectory is a spacetime geodesic. This definition is supported by the \textit{weak equivalence principle} [3], a result verified by multiple experiments, which establishes that all objects in freefall have the same acceleration, regardless of their composition or structure, thus revealing that the effects of gravity depend on a property that resides in gravity itself rather than in the objects affected by it. This property is the curvature of spacetime. In the following, we will use spacetime geodesics, together with the notion of geodesic deviation introduced in Part I, to obtain an intuitive picture of the curvature of spacetime.\\

As we know, the curvature of a manifold is determined from the geodesic deviation $\xi(t)$ of two initially parallel paths, whose magnitude decreases as a function of the parameter $t$ if the curvature is positive, increases if the curvature is negative, and remains constant if it is null\footnote{There is no universal convention for the assignment of "positive" and "negative" signs to spacetime curvature. In some textbooks the sign is chosen as in this paper, in others it is chosen with the opposite sign.} (holds under the condition that the two geodesics are initially parallel). In the context of GR we will interpret the parameter $t$ as the time that takes a particle to move freely through spacetime. Indeed, consider two test particles that fall freely from rest toward the centre of the Earth from a great height, as shown in Fig. 10. The particles are separated by a radial distance $\xi_{r}(t)$, and are inside a frame of reference \textit{S} that falls freely with them. We can imagine \textit{S} as a spaceship with its engines switched off, falling towards Earth.\\

\begin{figure}[h]
  \centering
    \includegraphics[width=0.55\textwidth]{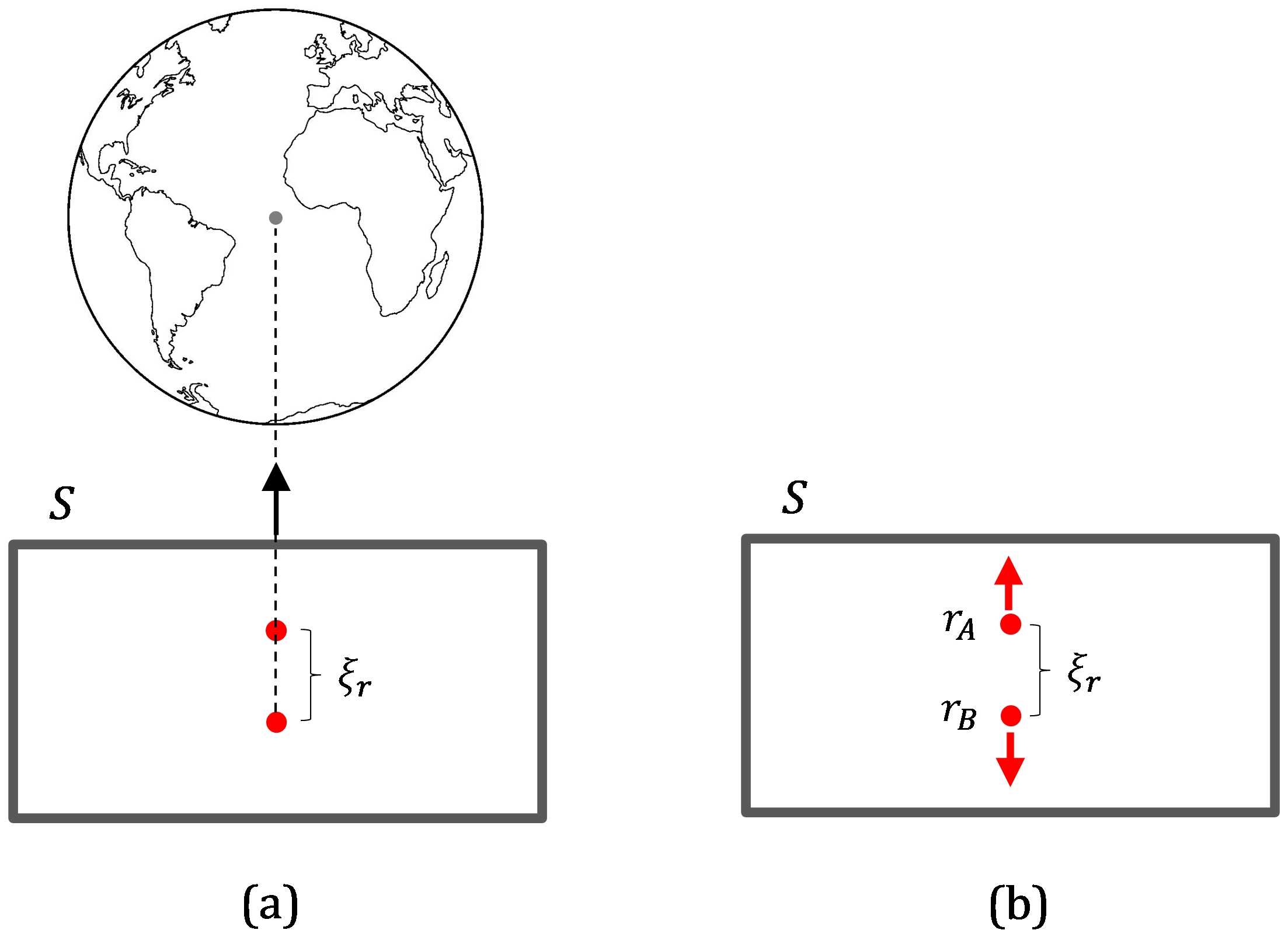}
  \caption{(a) \textit{S} falls freely together with two particles (shown in red); (b) inside \textit{S}, it is observed that $\xi_{r}(t)$ increases, and hence the space-time curvature is negative.}
\end{figure}

If $r_{A} (t)$ and $r_{B} (t)$ are the positions of the particles with respect to \textit{S}, then:

\begin{equation} 
\xi_{r}(t) = r_{A}(t) - r_{B}(t).
\end{equation}

The gravity acting on the particle closest to the Earth is greater than on the one farthest away, so that they separate with an accelerating motion, and in \textit{S} (and in all reference systems) it is observed that $\xi_{r}(t)$ increases (Fig. 1). Since the particles have an initial speed of zero, in a spacetime diagram the geodesics \textit{A} and \textit{B} are initially parallel but then diverge (Fig. 2), similarly to the surface of a saddle, which implies that the curvature of spacetime in the radial direction is negative.\\

\begin{figure}[h]
  \centering
    \includegraphics[width=0.6\textwidth]{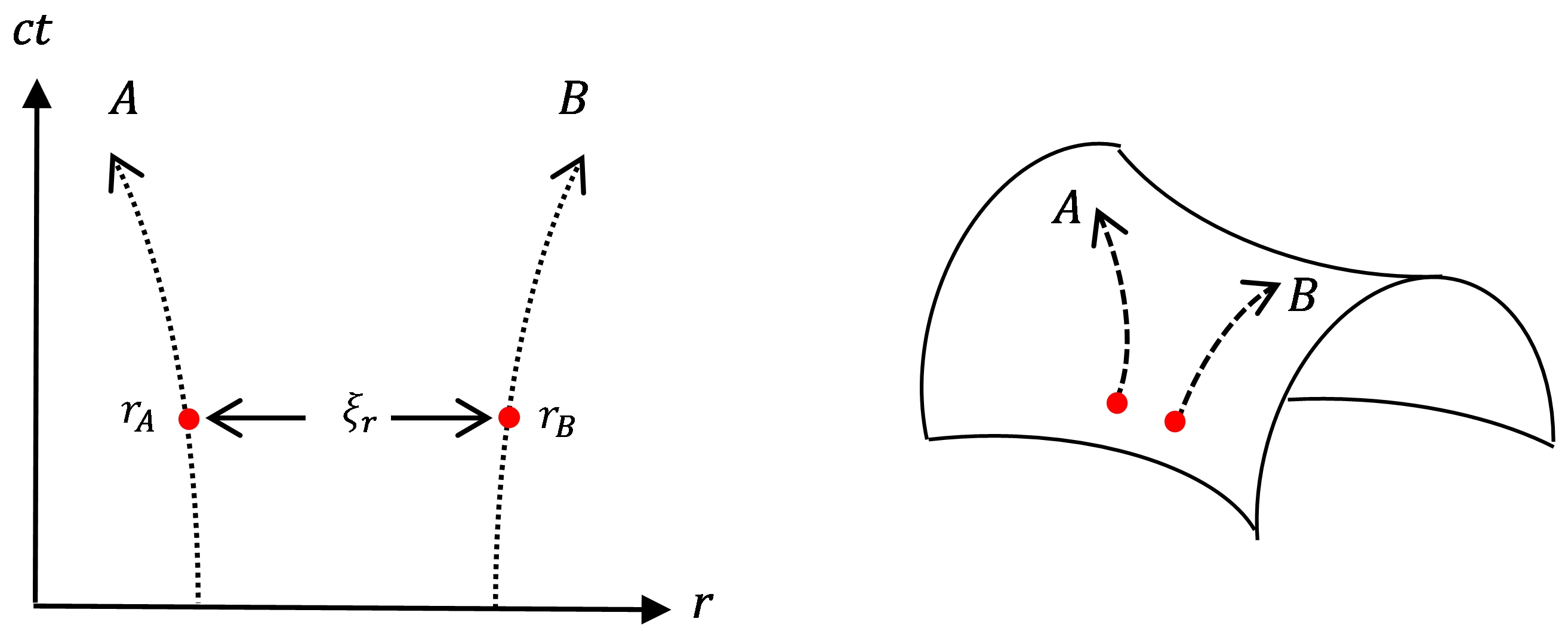}
  \caption{Geodesics \textit{A} and \textit{B} are initially parallel but then diverge, as on the surface of a saddle.}
\end{figure}

Suppose now that the particles are separated by a distance $\xi_{\theta}(t)$ along the direction transverse to the radius (Fig. 3). If $\theta_{A}(t)$ and $\theta_{B}(t)$ are the positions of the particles with respect to \textit{S}, we have

\begin{equation} 
\xi_{\theta}(t) = \theta_{A}(t) - \theta_{B}(t).
\end{equation}

The particles fall towards the centre of the Earth, so that they approach with an accelerating movement, and in \textit{S} (and in all other reference systems) it is observed that $\xi_{\theta}(t)$ decreases (Fig. 3). In a spacetime diagram, the geodesics \textit{A} and \textit{B} are initially parallel but then converge (Fig. 4) as on the surface of a sphere, which implies that the curvature of spacetime in the transverse direction is positive.\\

\begin{figure}[h]
  \centering
    \includegraphics[width=0.55\textwidth]{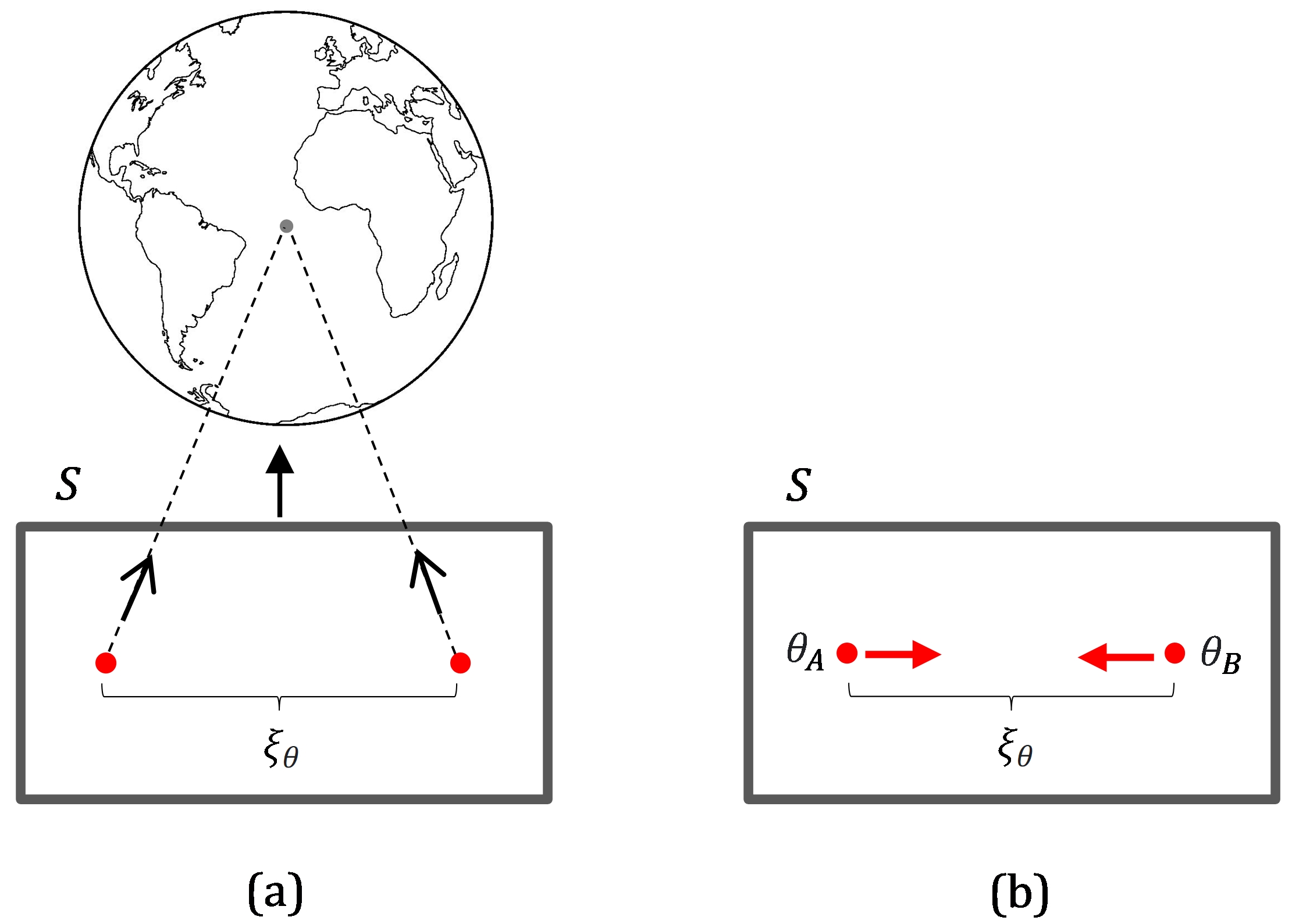}
  \caption{(a) \textit{S} falls freely together with two particles; (b) inside \textit{S} it is observed that $\xi_{\theta}(t)$ decreases, and therefore the curvature of spacetime is positive.}
\end{figure}

Let us now imagine that the particles are in a region of space where there is no gravity (Fig. 5). If the distance that separates them in \textit{S} coincides with the \textit{x} direction, then:

\begin{equation} 
\xi_{x}(t) = x_{A}(t) - x_{B}(t).
\end{equation}

If the particles are left at rest relative to each other and no force acts on them, in \textit{S} (and in all reference frames) it is observed that $\xi_{x}(t)$  is constant (Fig. 5). In a spacetime diagram, the geodesics \textit{A} and \textit{B} are parallel, as on the surface of a sheet of paper (Fig. 6), which implies that the curvature of spacetime is zero.\\

\begin{figure}[h]
  \centering
    \includegraphics[width=0.55\textwidth]{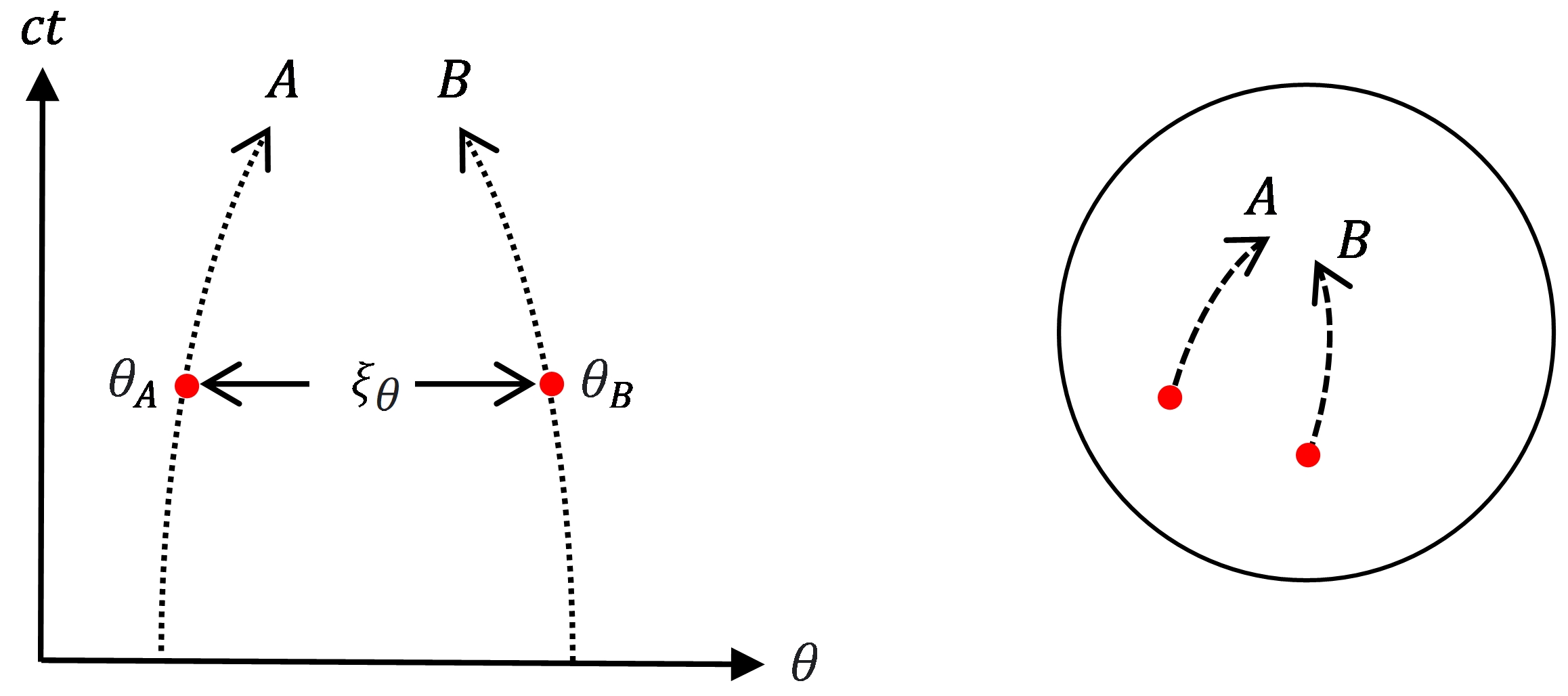}
  \caption{Geodesics \textit{A} and \textit{B} are initially parallel but then converge, as on the surface of a sphere.}
\end{figure}

To synthesise the previous ideas, since the geodesic deviation originates from gravity, we conclude that the spacetime curvature: (1) \textit{is caused by the mass}, and (2) \textit{depends on the orientation, that is, it is variable}. To these ideas we must add a relativistic peculiarity that we can only mention without proof: unlike the spatial geodesics of Riemannian geometry, which correspond to the shortest distance between two points, the spacetime geodesics correspond to the longest path; that is, of all the trajectories connecting two points in spacetime, a particle follows the one for which the proper time experienced by the particle is a maximum. But despite this difference between the Riemannian and spacetime manifolds, in both cases it is true that a geodesic is a locally straight line, which makes this definition more useful, as we pointed out in Part I.\\

\begin{figure}[h]
  \centering
    \includegraphics[width=0.25\textwidth]{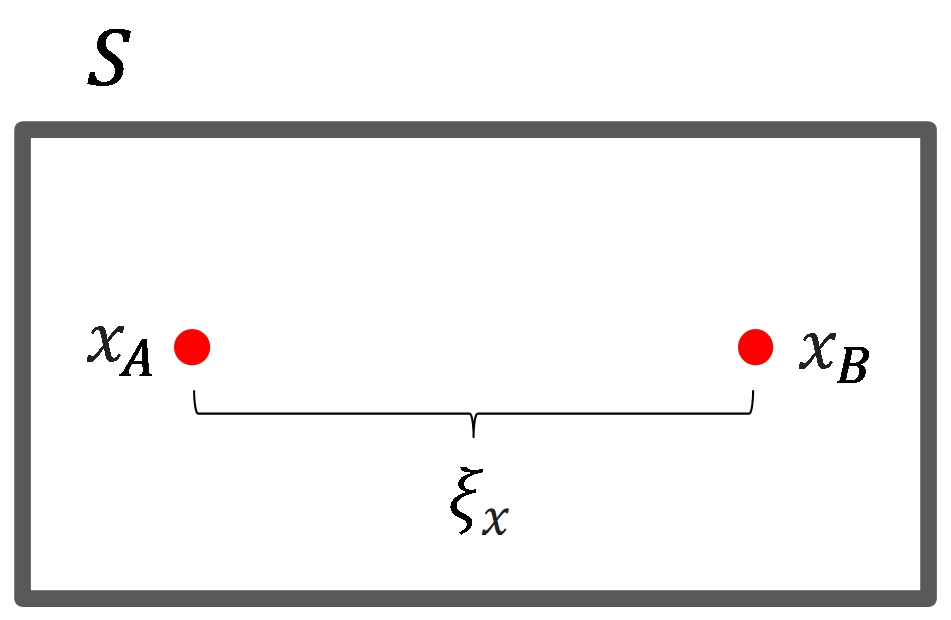}
  \caption{Inside \textit{S} it is observed that $\xi_{x}(t)$ remains constant, and hence the curvature of spacetime is null.}
\end{figure}

Let us now return to Figs. 1 and 3. If we consider infinitesimal time spans and assume that \textit{S} is very small, which is equivalent to considering a very small region of spacetime, $\xi_{r}(t)$ and $\xi_{\theta}(t)$ will remain approximately constant, and the corresponding geodesics will look almost straight and parallel, as if spacetime were flat. Here an important idea introduced in Part I reappears: \textit{a sufficiently small region of a curved manifold can be considered locally flat}. This also applies to spacetime geometry, and reinforces an idea introduced earlier: SR is locally valid in curved spacetime, and GR contains SR as a special case. But this statement presents a subtle physical problem, since SR is only strictly valid for inertial reference frames, where there is no gravity, and a particle falling freely towards a celestial body experiences gravity.\\

\begin{figure}[h]
  \centering
    \includegraphics[width=0.6\textwidth]{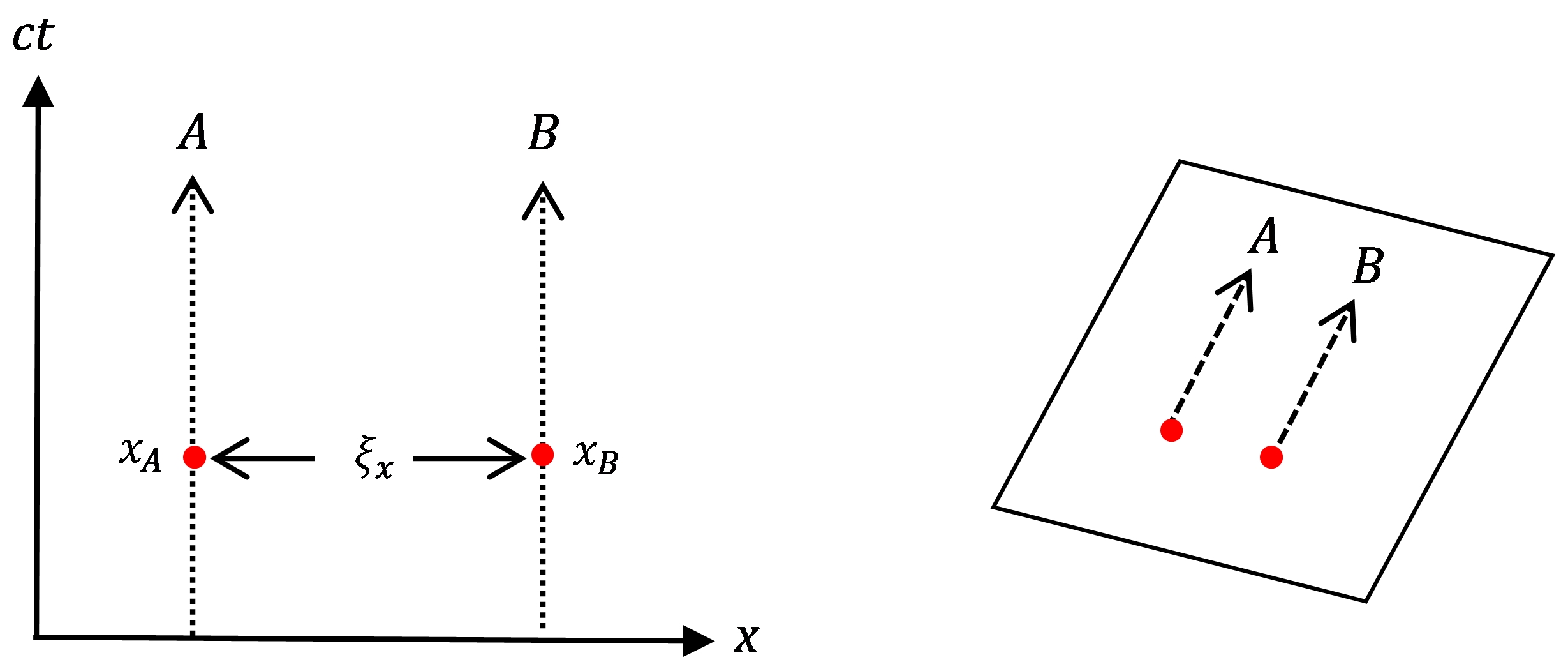}
  \caption{Geodesics \textit{A} and \textit{B} are parallel, as on the surface of a sheet of paper.}
\end{figure}

Einstein overcame this difficulty by noting that an observer in free fall feels weightless, as if he or she were floating in a region of space where there are no gravitating masses. The great physicist embodied this idea in his equivalence principle: \textit{A frame of reference in free fall in a gravitational field} (such as \textit{S} in Figs. 1 and 3) \textit{is locally equivalent to an inertial frame} (without gravity) [3,4]. Then, the equivalence principle guarantees that in a small neighbourhood of a curved spacetime point, SR agrees with GR.\\

Let us consider two important applications of this result. From Eqs. (32) and (33) of Part I, we know that in SR we have $ds = cd\tau$ for a material object and $ds=0$ for light; hence, these results must be valid in any small region of curved spacetime, so that Eq. (1) also satisfies these conditions.

\section{The Einstein equation}

The concepts of metric, curvature, and geodesic deviation discussed earlier provide the geometric framework for describing gravity. In particular, the ideas developed in the previous section give us some intuitive insights into the relationship between geometry and gravity. However, we still do not know the exact dynamics of the gravitational field, that is, the way in which the distribution of matter determines gravity. To achieve this, Einstein elaborated GR [4], which is expressed by a set of 16 equations that can be written compactly as the \textit{Einstein equation} [1,2]:

\begin{equation} 
G_{\mu \nu} = \frac{8\pi G}{c^{4}} T_{\mu \nu},
\end{equation}

or in matrix representation as,

\begin{equation} 
\begin{pmatrix} G_{00} & G_{01} & G_{02} & G_{03} \\ G_{10} & G_{11} & G_{12} & G_{13} \\ G_{20} & G_{21} & G_{22} & G_{23} \\ G_{30} & G_{31} & G_{32} & G_{33} \end{pmatrix} = \dfrac{8\pi G}{c^{4}} \begin{pmatrix} T_{00} & T_{01} & T_{02} & T_{03} \\ T_{10} & T_{11} & T_{12} & T_{13} \\ T_{20} & T_{21} & T_{22} & T_{23} \\ T_{30} & T_{31} & T_{32} & T_{33} \end{pmatrix},
\end{equation}

where $G$ is the gravitational constant, $c$ is the speed of light in vacuum, and the set of $G_{\mu \nu}$ and $T_{\mu \nu}$ define the \textit{Einstein tensor} and the \textit{energy-momentum tensor}, respectively. The Einstein tensor is a function of the metric tensor\footnote{For the expert reader, the Einstein tensor is defined as $G_{\mu \nu} \equiv R_{\mu \nu} -\frac{1}{2} Rg_{\mu \nu}$ where $R$ is the \textit{Ricci scalar}, and $R_{\mu \nu}$ is the \textit{Ricci tensor}. From the Einstein tensor, the Einstein equation is written as $R_{\mu \nu} -\frac{1}{2} Rg_{\mu \nu} = \frac{8\pi G}{c^{4}} T_{\mu \nu}$.}, which we can write as $G_{\mu \nu} = G_{\mu \nu}(g_{\mu \nu})$.\\

The Einstein equation relates $G_{\mu \nu} = G_{\mu \nu}(g_{\mu \nu})$, which describe the curvature at each point in spacetime, to $T_{\mu \nu}$, which describe the distribution of mass-energy that causes the curvature, and which are measured in units of energy density, that is, energy per unit of volume. The great physicist John Archibald Wheeler beautifully described the idea behind Einstein equation, noting that "mass tells spacetime how
to curve, and spacetime tells mass how to move" [5]. In symbolic form, Einstein equation is:

\begin{equation} 
\left(  \begin{array}{c} spacetime \\ curvature \end{array} \right) = \frac{8\pi G}{c^{4}}\left( \begin{array}{c} mass-energy \\ distribution  \end{array} \right).
\end{equation}

If we assume that the mass-energy distribution is zero, or if we consider the Newtonian limit at which there is no maximum speed of information transmission ($c \rightarrow \infty$), the right-hand side of this equality is zero and spacetime is flat:

\begin{equation} 
\left(  \begin{array}{c} spacetime \\ curvature \end{array} \right) = 0.
\end{equation}

In practice, the Einstein equation only contains ten independent equations, since the matrices are symmetric, so we can simplify the equation to:

\begin{equation} 
\begin{pmatrix} G_{00} & G_{01} & G_{02} & G_{03} \\ 0 & G_{11} & G_{12} & G_{13} \\ 0 & 0 & G_{22} & G_{23} \\ 0 & 0 & 0 & G_{33} \end{pmatrix} = \dfrac{8\pi G}{c^{4}} \begin{pmatrix} T_{00} & T_{01} & T_{02} & T_{03} \\ 0 & T_{11} & T_{12} & T_{13} \\ 0 & 0 & T_{22} & T_{23} \\ 0 & 0 & 0 & T_{33} \end{pmatrix}.
\end{equation}

Each of these ten equations is obtained by equating the components with the same subscripts, for example:

\begin{equation} 
G_{0 0} = \frac{8\pi G}{c^{4}} T_{0 0}, \ G_{0 2} = \frac{8\pi G}{c^{4}} T_{0 2}, \ G_{2 3} = \frac{8\pi G}{c^{4}} T_{2 3}.
\end{equation}

The central problem of GR is to solve the Einstein equation, that is, to find the values of $g_{\mu \nu}$ (contained in $G_{\mu \nu}$) for a given mass-energy distribution. In many situations of interest, $g_{\mu \nu}$ can be represented by a diagonal matrix, and the solutions to the Einstein equation have the form:

\begin{equation} 
\left[ g_{\mu \nu} \right]  = \begin{pmatrix} g_{00} & 0 & 0 & 0 \\ 0 & g_{11} & 0 & 0 \\ 0 & 0 & g_{22} & 0 \\ 0 & 0 & 0 & g_{33} \end{pmatrix},
\end{equation}

which in generalised coordinates $x_{0}, x_{1}, x_{2}, x_{3}$ has associated the metric

\begin{equation} 
ds^{2} = g_{00}dx_{0}^{2} + g_{11}dx_{1}^{2} + g_{22}dx_{2}^{2} + g_{33}dx_{3}^{2}.
\end{equation}

The term of the metric that contains the $x_{0}$ coordinate is called the \textit{temporal part}, and the terms that contain the coordinates $x_{1}, x_{2}, x_{3}$ are called the \textit{spatial part}. An example is the solution for a flat spacetime, devoid of matter, which in coordinates $x_{0} = ct, x_{1} =x, x_{2} = y, x_{3} = z$ is:

\begin{equation} 
\left[ g_{\mu \nu} \right]  = \left[ \eta_{\mu \nu} \right] = \begin{pmatrix} 1 & 0 & 0 & 0 \\ 0 & -1 & 0 & 0 \\ 0 & 0 & -1 & 0 \\ 0 & 0 & 0 & -1 \end{pmatrix},
\end{equation}

where, as we know from Eq. (28) of Part I:

\begin{equation} 
ds^{2} = c^{2}dt^{2} - dx^{2} - dy^{2} - dz^{2}.
\end{equation}

Another important example is the Schwarzschild metric, which we will discuss next.

\section{The Schwarzschild solution}

Many exact solutions to the Einstein equation are known, but only a few are of physical interest. This is the case for the first exact solution, found by astronomer Karl Schwarzschild in 1916, shortly after Einstein published his theory.\\

\begin{figure}[h]
  \centering
    \includegraphics[width=0.3\textwidth]{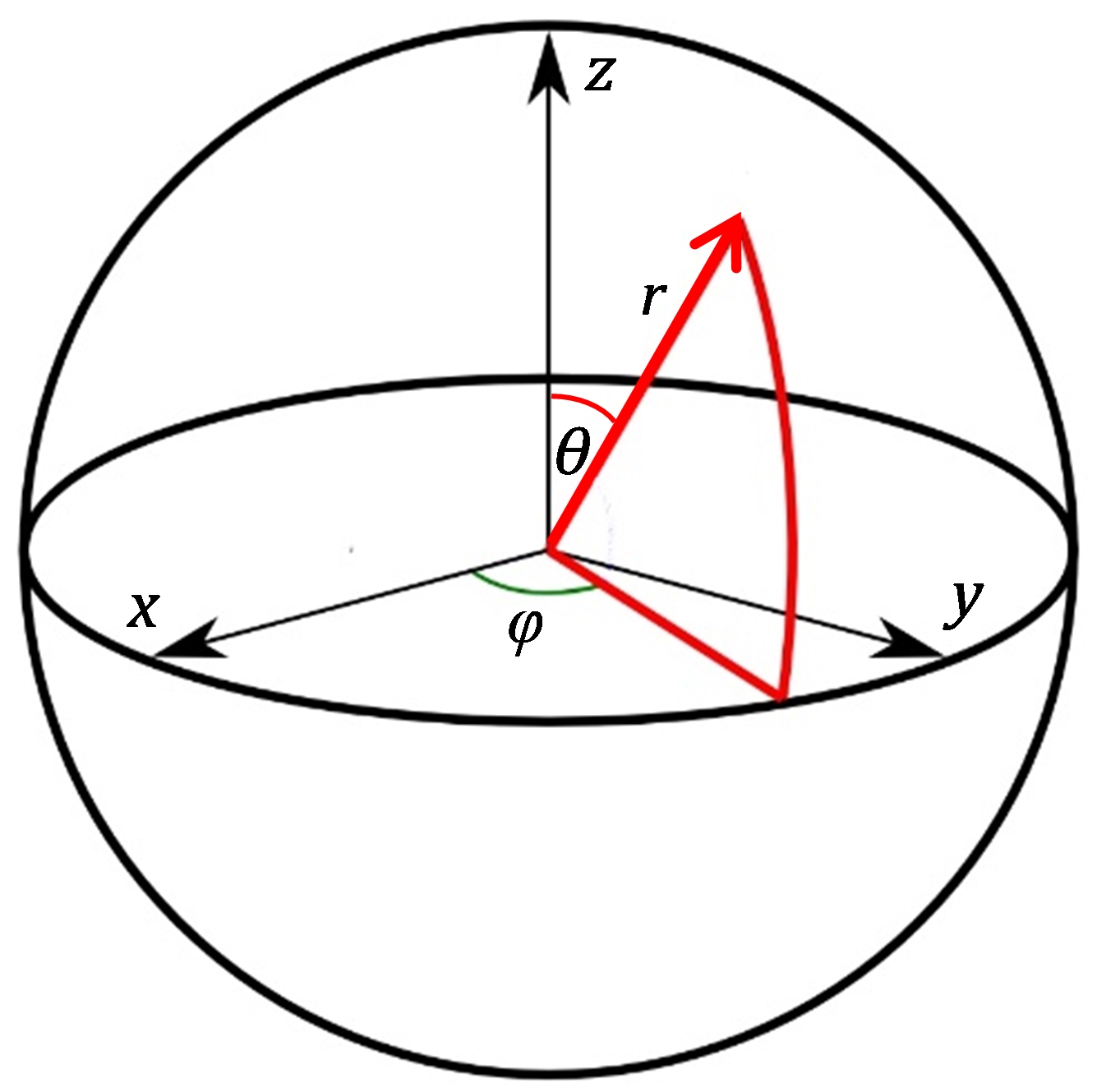}
  \caption{Spherical coordinates, $r > 0, 0 \leq \theta \leq \pi, 0 \leq \varphi \leq 2\pi$.}
\end{figure}

The Schwarzschild solution or \textit{Schwarzschild metric} (SM) describes the spacetime outside a spherically symmetric distribution of mass $M$ and radius $R$, similar to a planet or a star. This mass distribution is \textit{static} and \textit{stationary}, which means that it does not depend on time. Since outer spacetime is empty, Schwarzschild's finding is a vacuum solution. Given the mass distribution $M$, the problem that Schwarzschild faced is to determine $g_{\mu \nu} $ for $r > R$. Due to the symmetry of the problem, a first step is to use spherical coordinates $r,\theta, \varphi$, to which we add the time coordinate $t$, which allows us to define the \textit{Schwarzschild coordinates} (Fig. 7):

\begin{equation} 
x_{0} = ct, x_{1} = r, x_{2} = \theta, x_{3} = \varphi.
\end{equation}

In these coordinates, the solution found by Schwarzschild is [1,2]:

\begin{equation} 
\left[ g_{\mu \nu} \right]  = \begin{pmatrix} g_{00} & 0 & 0 & 0 \\ 0 & g_{11} & 0 & 0 \\ 0 & 0 & g_{22} & 0 \\ 0 & 0 & 0 & g_{33} \end{pmatrix} = 
\begin{pmatrix} 1-\frac{2GM}{c^{2}r} & 0 & 0 & 0 \\ 0 & -\left( 1-\frac{2GM}{c^{2}r} \right)^{-1}  & 0 & 0 \\ 0 & 0 & -r^{2} & 0 \\ 0 & 0 & 0 & -r^{2} \sin^{2}\theta \end{pmatrix}.
\end{equation} 

The SM can be written as [1,2]:

\begin{equation} 
ds^{2} = c^{2}\left( 1 - \frac{2GM}{c^{2}r} \right)dt^{2} - \left( 1 - \frac{2GM}{c^{2}r} \right)^{-1}dr^{2} - r^{2}d\theta^{2} - r^{2}\sin^{2}\theta d\phi^{2}.  
\end{equation}

As expected, this metric represents a static and stationary spacetime, which means that the Schwarzschild geometry does not experience changes in time. Mathematically this is reflected in the fact that the metric is invariant under the transformations $t\rightarrow -t$ and $t\rightarrow t+t_{0}$, where $t_{0}$ is a constant. The presence of coefficients that depend on $M$ suggests that this metric describes a curved spacetime. Eq. (18) can be written more compactly as:

\begin{equation} 
ds^{2} = c^{2}\left( 1 - \frac{R_{S}}{r} \right)dt^{2} - \left( 1 - \frac{R_{S}}{r} \right)^{-1}dr^{2} - r^{2}d\theta^{2} - r^{2}\sin^{2}\theta d\phi^{2}.  
\end{equation}

The quantity $R_{S}= 2GM/c^{2}$ is known as the \textit{Schwarzschild radius}, and is a measure of the size of a \textit{Schwarzschild black hole}, that is, a black hole with spherical symmetry, and which is static and stationary [1,6]. Each value of $M$ corresponds to a particular Schwarzschild radius, which represents the size to which an object of mass $M$ would have to be compressed in order for it to become a black hole. As an example, for the Earth we have $R_{S} \cong 10^{-2} m$, which is of the order of one hundred millionth of the Earth's radius, $R_{\oplus} = 6.5 \times 10^{6}m$.\\

Note that by taking $r \gg R_{S}$ ($R_{S}/r \ll 1$) in Eq. (19), it is reduced to Eq. (29) of Part I, which defines the Minkowski metric in spherical coordinates. This means that far from the Schwarzschild radius of a celestial body, gravity is so weak that the curvature is negligible and spacetime is almost Minkowskian, which suggests that Newtonian physics is a good approximation to describe gravity. When this happens, spacetime is said to be \textit{asymptotically flat}. This result suggests that the $R_{S} /r$ ratio is an indicator of the intensity of the GR effects. For example, in the case of the Earth, $R_{S}/ R_{\oplus} = 10^{-2} m/6.5 \times 10^{6} m \sim 10^{-8} \ll 1$, and hence the gravity at the Earth's surface is almost Newtonian. If we take $M = 0$ in Eq. (19), it is also reduced to the Minkowski metric in spherical coordinates, but unlike the previous situation, there is now no gravity and spacetime is completely flat.\\

Before finishing this section, it is worth making a technical clarification. To rigorously define the radial coordinate $r$ that appears in Eq. (16) it is necessary to use measurable quantities, but the radial coordinate $r$ is not measurable. However, the circumference $C$ surrounding the spherical mass distribution is measurable, so the radial coordinate is defined as $r= C/2\pi$. This will become clearer in Section 7 where we will introduce a graphical representation of Schwarzschild space-time called Flamm's paraboloid.\\

It is important to keep in mind that the information contained in the SM, and in general in any metric, is extracted by manipulating it mathematically, as exemplified in the Appendix 1 and 2.

\section{The Newtonian limit of the Einstein equation}

It can be shown that, when gravity is weak and spacetime curvature is small, the most important of the ten equalities contained in Eq. (7) is the one associated with time ($\mu, \nu = 0$), and we only need the other nine to describe extreme phenomena such as black holes. Then, the equation that we are interested in solving in the Newtonian limit is [1,2]:

\begin{equation} 
G_{0 0} = \frac{8\pi G}{c^{4}} T_{0 0}.
\end{equation}

We will use this result, together with the SM, to check heuristically that when gravity is weak, Einstein equation is reduced to Poisson equation,

\begin{equation}
\nabla^{2} \phi = 4\pi G\rho,
\end{equation}

which is the most general form that Newton's law of gravitation can take, where $\phi$ is the gravitational potential, $\rho$ is the mass density, and $\nabla^{2}$ is the \textit{Laplacian operator}\footnote{Recall that in Cartesian coordinates: $\nabla^{2} \equiv \frac{\partial^{2}}{\partial x^{2}} + \frac{\partial^{2}}{\partial y^{2}} + \frac{\partial^{2}}{\partial z^{2}}$}. This equation is the Newtonian equivalent of the Einstein equation, and relates gravity, described by $\nabla^{2} \phi$, to the mass density caused by gravity, described by $\rho$. Then, $\nabla^{2} \phi$ is the Newtonian equivalent of $G_{00}=G_{00}(g_{00})$, that depends of $g_{00}$, and $\rho$ is the equivalent of $T_{00}$ (Fig. 8). With these equivalences in mind, let us remember that the components $T_{\mu \nu}$ have units of energy density, so that in Eq. (20) we can take $T_{00} = \rho c^{2}$:

\begin{equation} 
G_{00} = \frac{8\pi G \rho}{c^{2}}.
\end{equation}

\begin{figure}[h]
  \centering
    \includegraphics[width=0.3\textwidth]{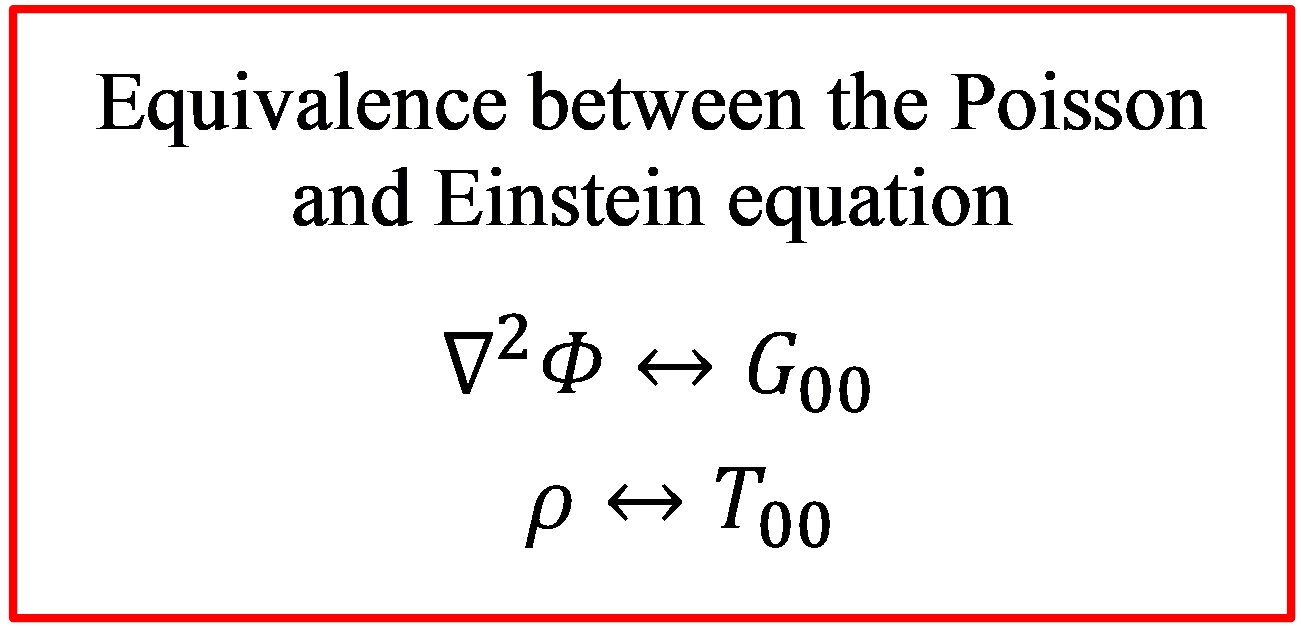}
  \caption{Formal equivalence between the Poisson equation, which governs Newtonian gravitation, and the Einstein equation, which governs relativistic gravitation.}
\end{figure}

On the other hand, $\phi$ is the Newtonian equivalent of $g_{00}-1$, which from Eq. (17) we can rewrite as:

\begin{equation} 
g_{00} -1 = \frac{2\phi}{c^{2}},
\end{equation}

where $\phi = -GM/r$ is the gravitational potential associated with a spherical celestial body of density $\rho$. We can then relate Eqs. (20) and (21) as follows:

\begin{equation} 
G_{00} = \nabla^{2} \left( g_{00}-1 \right) = \nabla^{2} \left( \frac{2\phi}{c^{2}} \right) = \frac{2}{c^{2}} \nabla^{2} \phi.
\end{equation}

By eliminating $G_{00}$ from Eqs. (22) and (24), we recover the Poisson equation.

\section{Flamm's paraboloid}

There is a powerful graphical representation of Schwarzschild geometry called \textit{Flamm's paraboloid}, which allows us to visualize the spacetime around a massive spherical object, such as a star or a black hole, by the trick of embedding said spacetime in a space of higher dimension [8,9]. Thus, we can visualize the curvature of a one-dimensional circumference, embedding it in a two-dimensional space. In the same way, we can visualize the curvature of a two-dimensional spherical surface, embedding it in a three-dimensional space. Let us apply these ideas to Schwarzschild spacetime.\\

\begin{figure}[h]
  \centering
    \includegraphics[width=0.9\textwidth]{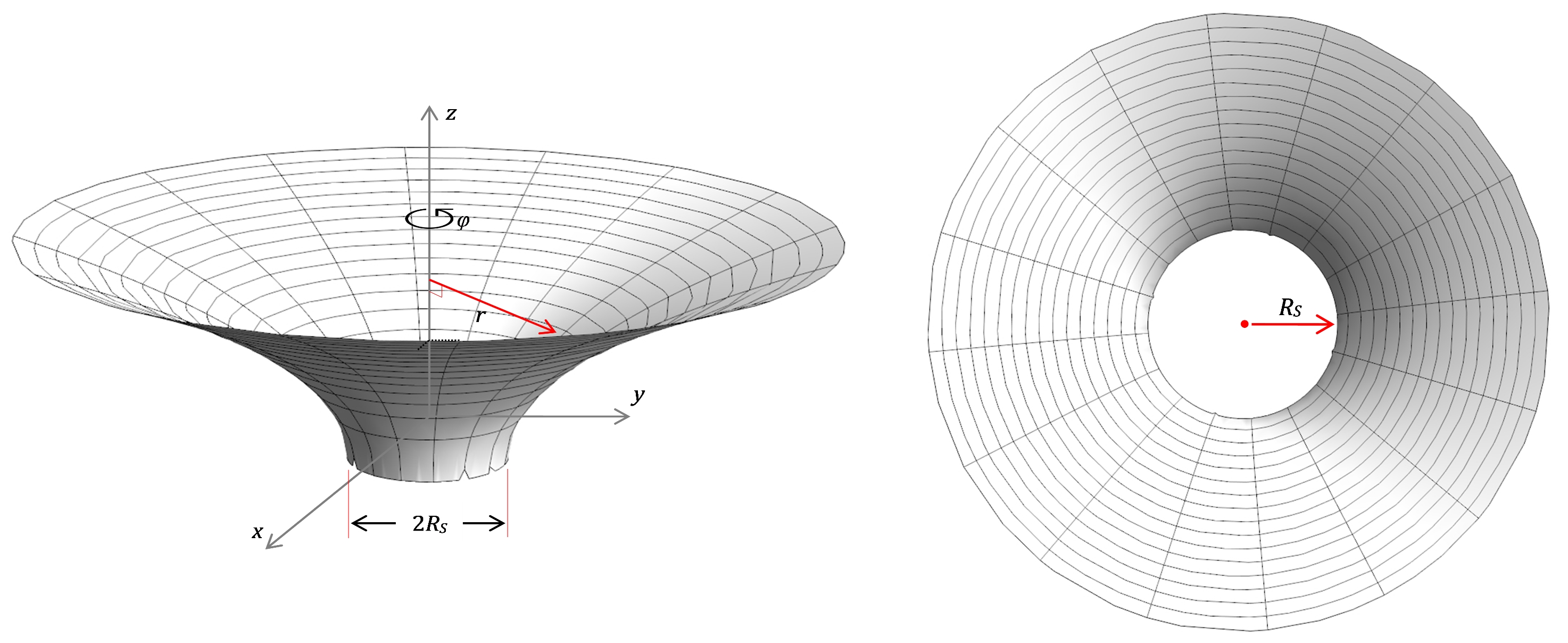}
  \caption{Side (left) and top (right) views of Flamm's paraboloid, representing Schwarzschild spacetime in two dimensions.}
\end{figure}

To do this, we must remember that the Schwarzschild geometry is static and stationary, which means that it is instructive to consider the three-dimensional space obtained by taking $t$ constant. This is equivalent to considering only the spatial part of the metric. Indeed, taking $t = constant$ in Eq. (19) and using the signature $(-,+,+,+)$ we obtain:

\begin{equation} 
ds^{2} = \left( 1 - \frac{R_{S}}{r} \right)^{-1}dr^{2} + r^{2} \left[ d\theta^{2} + \sin^{2}\theta d\phi^{2} \right].  
\end{equation}

Furthermore, since the Schwarzschild geometry is spherically symmetric, we can further simplify the description by analyzing the two-dimensional subspace obtained by taking $\theta = 90^{o} = constant$,

\begin{equation} 
ds^{2} = \left( 1 - \frac{R_{S}}{r} \right)^{-1}dr^{2} + r^{2}d\phi^{2}.  
\end{equation}

This expression describes the Schwarzschild geometry in the equatorial plane ($\theta = 90^{o}$) as a function of $r$. Thus, each value of $r$ and each rotation of $\phi$ by $2\pi$ ($360^{o}$) defines a circle centered on the $z$ axis. It can be shown that by continuously joining all these circles, a Flamm's paraboloid is obtained (see Fig. 9), which has cylindrical symmetry. The projection of the paraboloid on the plane $r-z$ defines a parabola. Therefore, we can also say that the continuous rotation of the parabola generates the paraboloid (Fig. 10). As can be shown from Eq. (19) (see Appendix 2), this parabola has the equation [8,9]:

\begin{equation}
z = \pm 2\sqrt{R_{S}(r-R_{S})} = \pm 2\sqrt{R_{S}(\sqrt{x^{2} + y^{2}}-R_{S})}. 
\end{equation}

This parabola intercepts the $r-axis$ at $R_{S}$; by taking $R_{S} = 1$, $r$ can be expressed in units of $R_{S}$. Fig. 10 shows the parabola in black, together with the same parabola rotated through $180^{o}$, shown in red. The solution for positive $z$ represents the spacetime of the known universe. There is also a solution for negative $z$ that does not appear in Fig. 9, and which is related to wormholes, a controversial topic that we will not analyze in this work.\\

In Fig. 9 the geometry inside $R_{S}$ does not appear, since for $r < R_{S}$, the value of $z$ in Eq. (27) is complex. The shape of the paraboloid is reminiscent of the rubber sheet analogy, used in elementary discussions to represent the curvature of spacetime. However, the similarity between both representations is superficial. The main difference is that only Flamm's paraboloid is mathematically rigorous. Furthermore, since Schwarzschild spacetime is static and stationary, the paraboloid is also stationary, which means that unlike a rubber sheet, there cannot be objects moving on the surface of the paraboloid. It can be said that the paraboloid is a sort of photograph of space-time.\\

\begin{figure}[h]
  \centering
    \includegraphics[width=0.7\textwidth]{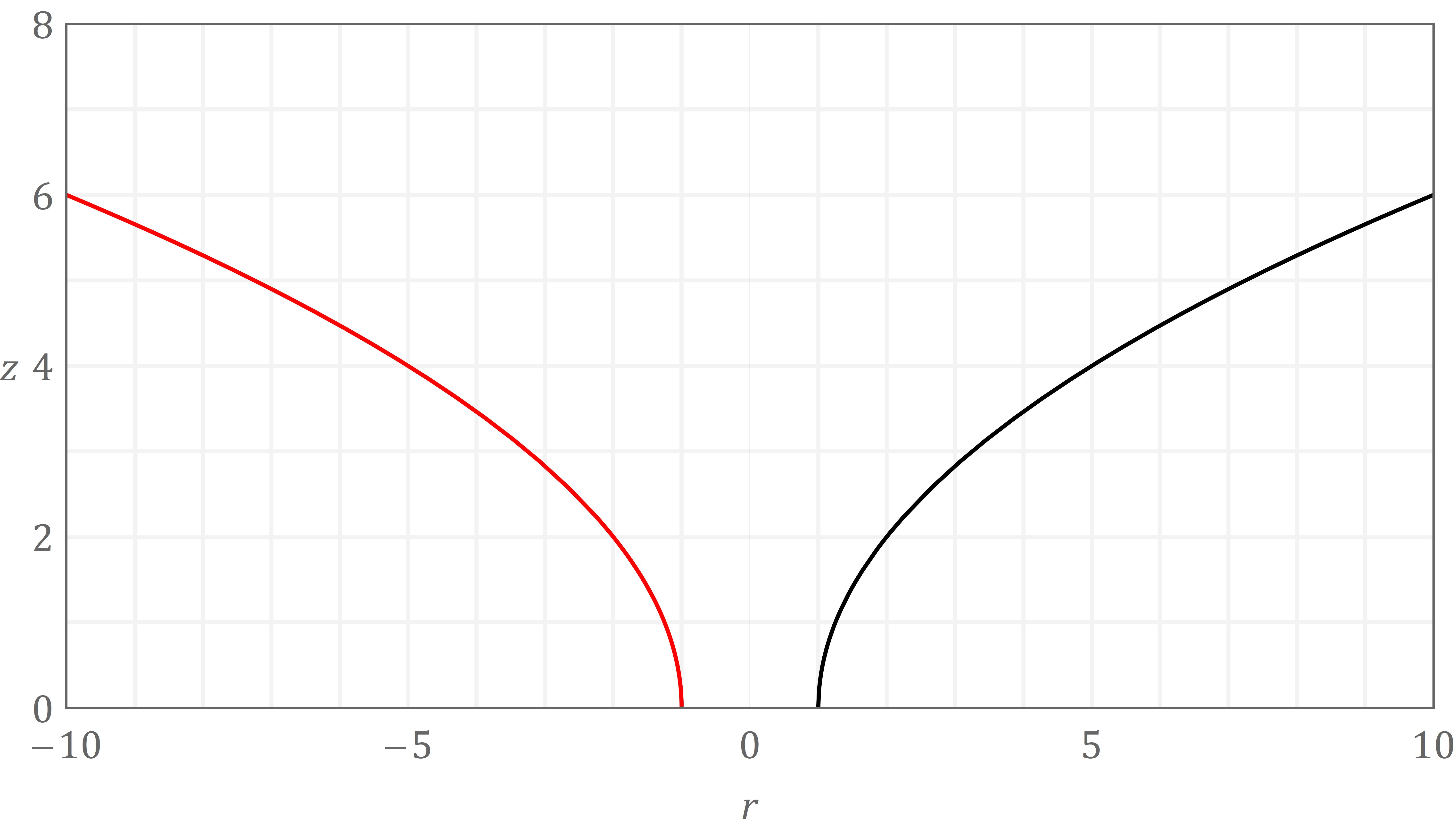}
  \caption{Parabola used to generate the two-dimensional Schwarzschild geometry.}
\end{figure}

It is important to consider that only the two-dimensional surface of the Flamm paraboloid has meaning as part of Schwarzschild geometry. Points outside the surface have no physical meaning. This allows us to understand intuitively the reason why we must define the radial coordinate by means of the equality $r=C/2\pi$, as we pointed out at the end of Section 5. Fig. 9 clearly shows us that the radius $r$ is not contained in the surface of the paraboloid, and therefore, that it lacks physical meaning. However, the circumferences $C$ are on the surface of the paraboloid and, consequently, they are measurable magnitudes.\\

If we want to know what spacetime outside an object of radius $R$ looks like, we just need to express $R$ in units of $R_{S}$ and then look at the paraboloid for $r \geq R$. For example, in the case of the Earth, $r = R_{\oplus} = 10^{8} R_{S} \gg R_{S}$, meaning that it is necessary to look very far from $r = R_{S}$. But, as Figs. 9 and 10 suggest, the surface of the paraboloid is \textit{asymptotically flat}, which implies that for $r = R_{\oplus}$, spacetime is almost flat. We can easily verify this by differentiating Eq. (27) to obtain the slope of the paraboloid surface in the radial direction:

\begin{equation}
\frac{dz}{dr} = \frac{R_{S}}{\sqrt{R_{S}(r-R_{S})}} =\frac{1}{\sqrt{r/R_{S}-1}}.
\end{equation}

The change in slope $dz/dr$ is a simple indicator of the curvature of spacetime in the radial direction. Thus, as we move away from the horizon, the slope decreases, and tends to zero ($(dz/dr) \approx 0$) when $r$ tends to infinity, which means that very far from the black hole ($r \gg R_{S}$), the curvature is almost zero. This conclusion is equivalent to the one we reached in Section 5, where we saw that outside the earth's surface, space-time is almost flat. The same holds for the other objects in the Solar system, including the Sun, and most of the stars in the universe.\\

This reveals that is it necessary to use GR only in extreme situations. However, these rare exceptions show us a world of wonders that have no place in Newtonian physics, such as black holes, neutron stars, gravitational lensing, active galaxies, and gravitational waves.

\section{Final comments}

GR has a reputation for being a very complex mathematical theory. This has greatly harmed the dissemination and teaching of GR, and has meant that only a few "brave" souls have dared to delve into its secrets. Although the breadth and complexity of the topic addressed here has forced us to be selective, it is hoped that this work will help improve the reputation of GR, thus contributing to its teaching and learning.

\section*{Appendix 1}

Let us consider a point in space located at a ﬁxed position with respect to a star of mass $M$. This implies that $dr = d\theta = d\varphi = 0$ in Eqs. (18) and (19). Taking $ds = cd\tau$, we have:

\begin{figure}[h]
  \centering
    \includegraphics[width=0.7\textwidth]{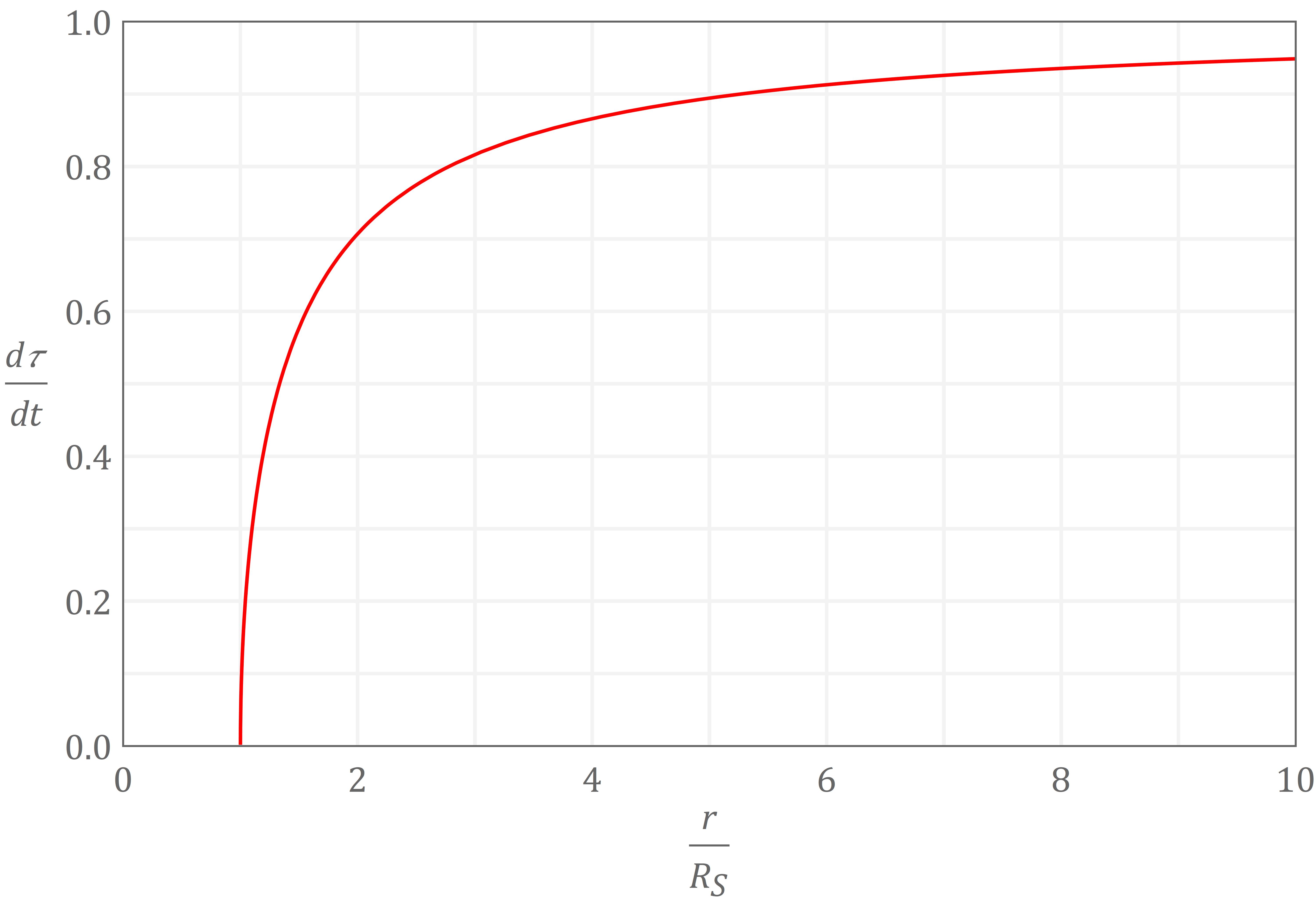}
  \caption*{Plot of $d\tau/dt$ versus $r/R_{S}$ illustrating the effect of gravitational time dilation.}
\end{figure}

\begin{equation} \tag{A1.1}
d\tau = \left( 1- \frac{2GM}{c^{2}r} \right)dt = \left( 1- \frac{R_{S}}{r} \right)dt, 
\end{equation}

where $d\tau$ is the proper time measured by a clock located at a distance $r$ from the gravitating body, and $dt$ is the coordinate time measured by a distant clock. We see that $d\tau < dt$, that is, the clock that measures proper time is behind the distant clock. This effect, known as \textit{gravitational time dilation}, intensifies with decreasing $r$ and reaches a maximum at $r = R_{S}$. To illustrate this effect, it is convenient to rewrite Eq. (A1) in the form:

\begin{equation} \tag{A1.2}
d\tau/dt  = \left( 1- \dfrac{1}{r/R_{S}} \right). 
\end{equation}

In the figure, we have plotted $d\tau/dt$ versus $r/R_{S}$. We see that for $r/R_{S} \gg 1$, $d\tau/dt \rightarrow 1$, but as $(r/R_{S}) \rightarrow 1$, $d\tau/dt \rightarrow 0$, that is, as we approach $R_{S}$, $d\tau$ becomes smaller and smaller compared to $dt$, and at $r =R_{S}$, $d\tau/dt = 0$, which means that according to a distant observer, time stops at $R_{S}$.

\section*{Appendix 2}
To obtain the Flamm paraboloid, it is necessary to embed the space-time described by Eq. (27) in a three-dimensional space, for which we use cylindrical coordinates:

\begin{equation} \tag{A2.1}
ds^{2} = dz^{2} + dr^{2} + r^{2} d\phi^{2}. 
\end{equation}

This equation can be rewritten as,

\begin{equation} \tag{A2.2}
ds^{2} = \left( \frac{dz}{dr} \right)^{2}dr^{2} + dr^{2} + r^{2} d\phi^{2} = \left[ \left( \frac{dz}{dr} \right)^{2} + 1 \right] dr^{2} + r^{2} d\phi^{2}.
\end{equation}

If the Eq. (27) and this equation represent the same geometry must hold that:

\begin{equation} \tag{A2.3}
 \left[ \left( \frac{dz}{dr} \right)^{2} + 1 \right] dr^{2} + r^{2} d\phi^{2} =\left( 1 - \frac{R_{S}}{r} \right)^{-1}dr^{2} + r^{2}d\phi^{2}.
\end{equation}

Then,

\begin{equation} \tag{A2.4}
\left( \frac{dz}{dr} \right)^{2} + 1 = \left( 1 - \frac{R_{S}}{r} \right)^{-1} \rightarrow \frac{dz}{dr} = \pm \left( \frac{r}{R_{S}} - 1 \right)^{-1/2}.
\end{equation}

Integrating we recover Eq. (28):

\begin{equation} \tag{A2.5}
z = \pm \int_{0}^{r} \dfrac{dr}{\sqrt{\frac{r}{R_{S}} - 1}} = \pm 2\sqrt{R_{S}(r-R_{S})} = \pm 2\sqrt{R_{S}(\sqrt{x^{2} + y^{2}}-R_{S})}.
\end{equation}

\section*{Acknowledgments}
I would like to thank to Daniela Balieiro and Michael Van Sint Jan for their valuable comments in the writing of this paper. I would also like to thank the referees, whose comments and suggestions have allowed me to significantly improve this paper.

\section*{References}

[1] B. Schutz, A First Course in General Relativity, 2nd ed., Cambridge University Press, Cambridge, 2009.

\vspace{2mm}

[2] R. Lambourne, Relativity, Gravitation and Cosmology, Cambridge University Press, Cambridge, 2010.

\vspace{2mm}

[3] J. Pinochet, Classical Tests of General Relativity Part I: Looking to the Past to Understand the Present, Physics Education. 55 (2020) 65016.

\vspace{2mm}

[4] A. Einstein, Die Grundlage der allgemeinen Relativitätstheorie, Annalen Der Physik. 354 (1916) 769–822.

\vspace{2mm}

[5] J.A. Wheeler, A journey into gravity and spacetime, W. H. Freeman and Company, New York, 1990.

\vspace{2mm}

[6] J. Pinochet, Five misconceptions about black holes, Phys. Educ. 54 (2019) 55003.

\vspace{2mm}

[7] T.P. Cheng, A College Course on Relativity and Cosmology, Oxford University Press, Oxford, 2015.

\vspace{2mm}

[8] W. Rindler, Relativity: Special, General and Cosmological, 2nd ed., Oxford University Press, New York, 2006.

\vspace{2mm}

[9] A.M. Steane, Relativity Made Relatively Easy, Oxford University Press, Oxford, 2012.

\end{document}